\documentclass[a4paper,10pt]{article}
\usepackage{color}
\usepackage{amsmath}
\usepackage{graphicx}
\title{A reassessment of  Leggett inequality}
\author{Antonio Di Lorenzo\\ Instituto de F\'{\i}sica, Universidade Federal de Uberl\^{a}ndia, \\
38400-902 Uberl\^{a}ndia, Minas Gerais, Brazil}
\date{April 9, 2011}
\begin{document}
%\citeindextrue
\maketitle
\begin{abstract}
Leggett formulated an inequality which seems to generalize the Bell theorem to nonlocal hidden variable theories. 
Leggett inequality is violated by quantum mechanics, as was confirmed by experiment. 
However, a careful analysis reveals that the theory applies to a class of local theory. 
Contrary to what happens in the derivation of Bell inequality, it is not necessary 
to make the hypothesis of outcome independence to derive the Leggett inequality. 
\end{abstract}

\section{Introduction}

A hidden-variable model is a theory making use of a set of parameters which complement or replace altogether 
the quantum mechanical state in the determination of the probability of experimental outcomes 
and which are distributed in such a way 
that the average observed frequencies reproduce the experimental results with an accuracy comparable 
to that of quantum mechanics. A hidden-variable model may be experimentally testable if it makes 
predictions different from quantum mechanics, or, in the case it reproduce quantum mechanics on average, 
if it provides a protocol to measure or fix the additional parameters, 
so that the experiments can access the conditional 
probability $P(\text{event}|\text{parameters})$. 

In the beginning, the formulation of hidden-variable models was perhaps driven by the wish to find 
a more intuitive theory than quantum mechanics. In the present author's opinion, searching for a 
more fundamental theory is a worthy enterprise in itself, 
independently of whether it provide an intuitive account of the phenomena or not. Indeed, it would be arrogant and shortsighted to think that quantum mechanics is the ultimate theory and that it will never be replaced. 
In this sense, quantum mechanics is incomplete as any other physical theory and as our quest for understanding 
Nature. 

The controversial paper by Einstein, Podolsky, and Rosen \cite{Einstein1935} argued that, 
assuming locality holds, quantum mechanics is an incomplete theory, according to some arbitrary criteria
of reality and completeness formulated therein (however, see Refs.~\cite{Ruark1935,Bohr1935,Kemble1935,Furry1936a,Wolfe1936} for criticisms). 
After Bohm \cite{Bohm1952a,Bohm1952b} provided an example of a nonlocal theory which, 
by construction, reproduces quantum mechanics, 
the question of whether quantum mechanics could be reproduced or at least approximated 
by a local theory became relevant. 
The reformulation of the EPR argument in terms of spin \cite{Bohm1957} 
paved the way to the formulation of Bell inequality \cite{Bell1964}. 
This inequality holds under the hypotheses that (i) hidden variables exist, (ii) locality holds, 
(iii) outcome independence is satisfied (i.e. the conditional probability of observing an event $e$ 
for given values of the additional parameters and for given events at space-like separated regions 
is equal to the marginal probability for observing $e$).  
Bell proved that a judicious combination of the spin correlators 
$C_{\Psi,D}=\sum_{\sigma,\tau} \sigma \tau P_{\Psi,D}(\sigma,\tau)$ 
obtained by varying the settings $D$, which are given by the orientations $\mathbf{a}$ and $\mathbf{b}$ of two spin detectors, for a fixed preparation $\Psi$ of a system, satisfies an inequality.\footnote{
Actually, Ref.~\cite{Bell1964} assumed deterministic theories, so that hypothesis (iii) was 
 replaced by the stricter hypothesis of determinism, while stochastic theories 
obeying condition (iii) were treated 
in Refs.~\cite{Bell1971,Clauser1974}. 
Condition (iii), however, was assumed implicitly, and it was first pointed out in Ref.~\cite{Jarrett1984}, 
which referred to it as ``completeness'', while the more neutral and technical term ``outcome independence'' 
was coined by Shimony \cite{Shimony1990}.} 
This inequality is violated by experiments \cite{Freedman1972,Aspect1982a,Aspect1982b,Tapster1994,Tittel1998,Weihs1998}, 
which agree well with quantum mechanics. 
Thus, one of the hypotheses (i), (ii), or (iii) must be false. 
	
By the end of the 70s Leggett worked out another inequality, which was only published a 
quarter of a century later \cite{Leggett2003}. 
According to Ref.~\cite{Leggett2003} the inequality applies to a class of nonlocal theories 
which satisfy determinism and outcome independence. As Bell inequality, Leggett inequality is violated by quantum mechanics and, more importantly, disproved by experimental data \cite{Groblacher2007,Paterek2007,Branciard2007,Eisaman2008,Romero2010,Paternostro2010,Lee2011}. 
Furthermore, the bonds on Leggett inequality were improved \cite{Branciard2008}. 
A criticism about the assumptions underlying Leggett inequality and the significance of the experiments was presented in Ref. \cite{Suarez2008}, but it failed to address the issue of locality.

Here we prove that Leggett inequality, even in the stricter formulation of Ref. \cite{Branciard2008}, 
actually applies to a family of \emph{local stochastic} hidden-variable theories 
which do not necessarily satisfy the hypothesis of outcome independence. Hence, the significance of the 
experimental violation of Leggett inequality is to be reassessed:  no rejection of the so-called ``nonlocal realism'' 
\cite{Groblacher2007,Paterek2007,Branciard2007,Eisaman2008,Romero2010,Paternostro2010,Lee2011} 
is supported by the experiments. However, the fact that the theories satisfying Leggett inequality (and thus 
excluded by the experiments) do not require the hypothesis of outcome independence is of great relevance, since 
we can thus exclude a class of local theories not contained in the class satisfying Bell inequality, 
which was already ruled out by experiment \cite{Freedman1972,Aspect1982a,Aspect1982b,Tapster1994,Tittel1998,Weihs1998}. 

\section{The hypotheses of Leggett}
Let us  analyse the hypotheses at the basis of Leggett inequality. 
Ref.~\cite{Leggett2003} assumes a set of deterministic hidden variables formed by two unit vectors and additional parameters,  so that 

{\bf Hypothesis (A):}
$\Lambda=\{\mathbf{u},\mathbf{v},\lambda\}$ represents the hidden variables. 
The observable joint probability of observing outcomes $\sigma$ and $\tau$ 
in a EPR-Bohm experiment can then be written 
\begin{equation}\label{eq:probhidlegg}
P(\sigma,\tau|\mathbf{a},\mathbf{b}) = \int d\mu_{\mathbf{a},\mathbf{b}}(\Lambda) 
P(\sigma,\tau|\Lambda;\mathbf{a},\mathbf{b})\ , 
\end{equation}
with $d\mu_{\mathbf{a},\mathbf{b}}(\Lambda)$ an invariant measure of the 
hidden variables and $P(\sigma,\tau|\Lambda;\mathbf{a},\mathbf{b})$ 
the probability of obtaining $\sigma$ and $\tau$ for fixed $\Lambda$. 

{\bf Hypothesis (B) (locality of the distribution):}
The probability distribution for the parameters is assumed to be independent 
of the settings of the detectors and to factorize as
$\rho(\Lambda)=F(\mathbf{u},\mathbf{v})g_{\mathbf{u},\mathbf{v}}(\lambda)$.  
Let us call $\Delta_{\mathbf{u},\mathbf{v}}$ the domain of $\lambda$ for given values of 
$\mathbf{u}$ and $\mathbf{v}$. 

{\bf Hypothesis (C) (determinism):}
For given $\Lambda$, the outcome of either measurement is predictable with certainty, i.e., 
there exist two functions $A$ and $B$ such that 
$\sigma=A(\Lambda;\mathbf{a},\mathbf{b})$ and $\tau=B(\Lambda;\mathbf{b},\mathbf{a})$. 
In other words, the joint probability $P(\sigma,\tau|\Lambda;\mathbf{a},\mathbf{b})$ takes the value one in a subdomain 
$\Delta_{\mathbf{u},\mathbf{v};\mathbf{a},\mathbf{b}}(\sigma,\tau) \subseteq \Delta_{\mathbf{u},\mathbf{v}}$ and 
zero in the complementary subdomain. 

{\bf Hypothesis (C$'$) (outcome independence):} if the outcome of one measurement is known, the outcome of the other is not influenced. 
This further hypothesis is unnecessary if hypothesis (C) is made, since it is 
actually a consequence of determinism: $\Lambda$ carries all the relevant information 
to determine the outcome of a measurement, hence any additional information, like the outcome of another measurement, is redundant. 
In terms of probabilities, hypothesis (C$'$) implies that 
\begin{equation}
\label{eq:factor}
P(\sigma,\tau|\Lambda;\mathbf{a},\mathbf{b})=P(\sigma|\Lambda;\mathbf{a},\mathbf{b}) P(\tau|\Lambda;\mathbf{b},\mathbf{a})\ .
\end{equation} 
If hypothesis (C) is assumed, Eq. \eqref{eq:factor} can be restated equivalently saying that 
$\Delta_{\mathbf{u},\mathbf{v};\mathbf{a},\mathbf{b}}(\sigma,\tau)=
\Delta_{\mathbf{u},\mathbf{v};\mathbf{a},\mathbf{b}}(\sigma)\cap \Delta_{\mathbf{u},\mathbf{v};\mathbf{b},\mathbf{a}}(\tau)$, where $\Delta_{\mathbf{u},\mathbf{v};\mathbf{a},\mathbf{b}}(\sigma)$ represents the subdomain of $\Delta_{\mathbf{u},\mathbf{v}}$ 
in which $P(\sigma|\Lambda;\mathbf{a},\mathbf{b})=1$ and $\Delta_{\mathbf{u},\mathbf{v};\mathbf{b},\mathbf{a}}(\tau)$ 
is defined analogously. 
However, both hypotheses (C) and (C') will be shown later on to be unnecessary. 

{\bf Hypothesis (D) (compliance with Malus's law):} The marginal probability of observing the outcome $\sigma$ for fixed $\mathbf{u},\mathbf{v}$ 
should satisfy (we use the formalism for spin, not for polarization of light)
\begin{align}
%\nonumber
P(\sigma|\mathbf{u},\mathbf{v};\mathbf{a},\mathbf{b})\equiv& \sum_\tau 
P(\sigma,\tau|\mathbf{u},\mathbf{v};\mathbf{a},\mathbf{b}) 
%\\
\label{eq:malus}
%=&
 =\frac{1}{2}\left(1+\sigma \mathbf{u}\cdot\mathbf{a}\right) \ .
\end{align}

We remark that locality was not assumed for the probabilities $P(\sigma|\Lambda;\mathbf{a},\mathbf{b})$ and 
$P(\tau|\Lambda;\mathbf{b},\mathbf{a})$, since each of them can depend on the setting of the remote detector. 
However,  a local (i.e. independent of the detector settings) 
distribution $d\mu(\Lambda)$ for the  hidden variables was assumed. 
Thus it seems that locality and nonlocality hypotheses are made at the same time. 
However, we notice that the nonlocality depends on the level of the description. 
In terms of the variables $\mathbf{u}$ and $\mathbf{v}$ the theory is local, since the marginal probabilities are by hypothesis (D) independent of the setting of the far-away detector, and by hypothesis (B) the distribution 
of $\mathbf{u}$ and $\mathbf{v}$ is independent of the settings of the detectors. 

\section{Sufficient hypotheses for the validity of Leggett inequality}\label{sec:righthyps}
In the preceding section, we exposed the hypotheses made in Ref.~\cite{Leggett2003}. 
Now, we formulate an alternative, less restrictive set of hypotheses, and we show that Leggett inequality follows from them.
They are as follows:

{\bf Hypothesis 1:} the hidden variables are formed by two unit vectors 
$\mathbf{u}$ and $\mathbf{v}$, such that 
the observable joint probability of obtaining outcomes $\sigma$ and $\tau$ in the left and right arm of an EPR-Bohm 
setup is 
\begin{equation}\label{eq:probhid}
P(\sigma,\tau|\mathbf{a},\mathbf{b}) = \int dF_{\mathbf{a},\mathbf{b}}(\mathbf{u},\mathbf{v}) 
P(\sigma,\tau|\mathbf{u},\mathbf{v};\mathbf{a},\mathbf{b})\ , 
\end{equation}
where $dF$ is an invariant measure, which in a given coordinate system $\mathcal{S}$ takes the form 
$ dF_{\mathbf{a},\mathbf{b}}(\mathbf{u},\mathbf{v})=F_{\mathcal{S}}(x_j) dx_1 \cdots dx_n$. 
Hypothesis 1 does not mean that there can not be any more hidden variables, but that after integrating out all of them, except $\mathbf{u}$ and $\mathbf{v}$, Eq.~\eqref{eq:probhid} holds.  

{\bf Hypothesis 2 (locality):} (a) the measure $dF$ is local (i.e. independent of the detectors)  
$dF_{\mathbf{a},\mathbf{b}}(\mathbf{u},\mathbf{v})=dF(\mathbf{u},\mathbf{v})$
and (b) the marginal probability of obtaining outcome $\sigma$ on one arm does not 
depend on the setting of the detector in the other arm, namely 
\[P(\sigma|\mathbf{u},\mathbf{v};\mathbf{a},\mathbf{b})\equiv \sum_\tau 
P(\sigma,\tau|\mathbf{u},\mathbf{v};\mathbf{a},\mathbf{b}) =P(\sigma|\mathbf{u},\mathbf{v};\mathbf{a})  \ .
\]

{\bf Hypothesis 3 (compliance with Malus's law):} identical to hypothesis (D) above.

No more hypotheses are needed. If there are additional hidden variables $\lambda$ no assumption
 is needed about the form 
of the probability $P(\sigma,\tau|\mathbf{u},\mathbf{v},\lambda;\mathbf{a},\mathbf{b})$ nor about 
the distribution $d\mu_{\mathbf{a},\mathbf{b}}(\mathbf{u},\mathbf{v},\lambda)$, provided that 
Hypothesis 2 and 3 are satisfied. 
In particular, the hypothesis of outcome independence is unnecessary, contrary to what happens 
in the derivation of Bell inequality.
Here we rederive the inequality following 
Refs. \cite{Leggett2003} and \cite{Groblacher2007b}. 
The detectors are characterized by two unit vectors $\mathbf{a},\mathbf{b}$. 
As stated in Ref.~\cite{Groblacher2007b}, the following equality holds for any two variables 
taking values $\pm 1$
\begin{equation}\label{eq:basiceq}
-1+\left|\sigma +\tau \right| = \sigma \tau 
= 1-\left|\sigma -\tau \right|   \  .
\end{equation}
After multiplying Eq.~\eqref{eq:basiceq} by the probability 
$P(\sigma,\tau|\mathbf{u},\mathbf{v};\mathbf{a},\mathbf{b})$, we sum over $\sigma$ and $\tau$
 and we exploit the well known inequality 
$\overline{|x|}\ge\left|\overline{x}\right|$, so that 
we have, after applying hypothesis 3, 
\begin{align}\label{eq:basicineq}
-1+\left|\mathbf{u}\cdot\mathbf{a} +\mathbf{v}\cdot\mathbf{b} \right|
\le C(\mathbf{u},\mathbf{v};\mathbf{a},\mathbf{b}) \le 
1- \left|\mathbf{u}\cdot\mathbf{a}-\mathbf{v}\cdot\mathbf{b} \right| \ , 
\end{align} 
with the correlator $C(\mathbf{u},\mathbf{v};\mathbf{a},\mathbf{b})\equiv 
\sum_{\sigma,\tau} \sigma\tau P(\sigma,\tau|\mathbf{u},\mathbf{v};\mathbf{a},\mathbf{b})$. 
Upon averaging Eq.~\eqref{eq:basicineq} gives 
\begin{align}\label{eq:basicineq2}
-1+\int dF \left|\mathbf{u}\cdot\mathbf{a} +\mathbf{v}\cdot\mathbf{b} \right|
\le C(\mathbf{a},\mathbf{b}) \le 
1-\int dF \left|\mathbf{u}\cdot\mathbf{a}-\mathbf{v}\cdot\mathbf{b} \right| \ , 
\end{align} 
with $C(\mathbf{a},\mathbf{b})\equiv \sum_{\sigma,\tau} \sigma \tau P(\sigma,\tau|\mathbf{a},\mathbf{b})$ the spin correlator.\footnote{The hidden-variable theory might predict that 
the average value observed at one arm 
$\langle\sigma \rangle_{\mathbf{a},\mathbf{b}}\equiv\sum \sigma 
P(\sigma,\tau|\mathbf{a},\mathbf{b})\neq 0$. In this case $C(\mathbf{a},\mathbf{b})$ would not be the correlator, but simply the spin-spin average. However, 
$\langle\sigma \rangle_{\mathbf{a},\mathbf{b}}\neq 0$ 
would already contradict existing experiments.}
From here on, the derivation of the inequality is identical as that presented in 
Refs.~\cite{Leggett2003} and \cite{Groblacher2007b} and we reproduce it in the appendix, 
where we  evidentiate some fine points in the derivation. 

\section{Conclusions}
\begin{center}
\begin{figure}
\centering
\includegraphics[width=3.3in]{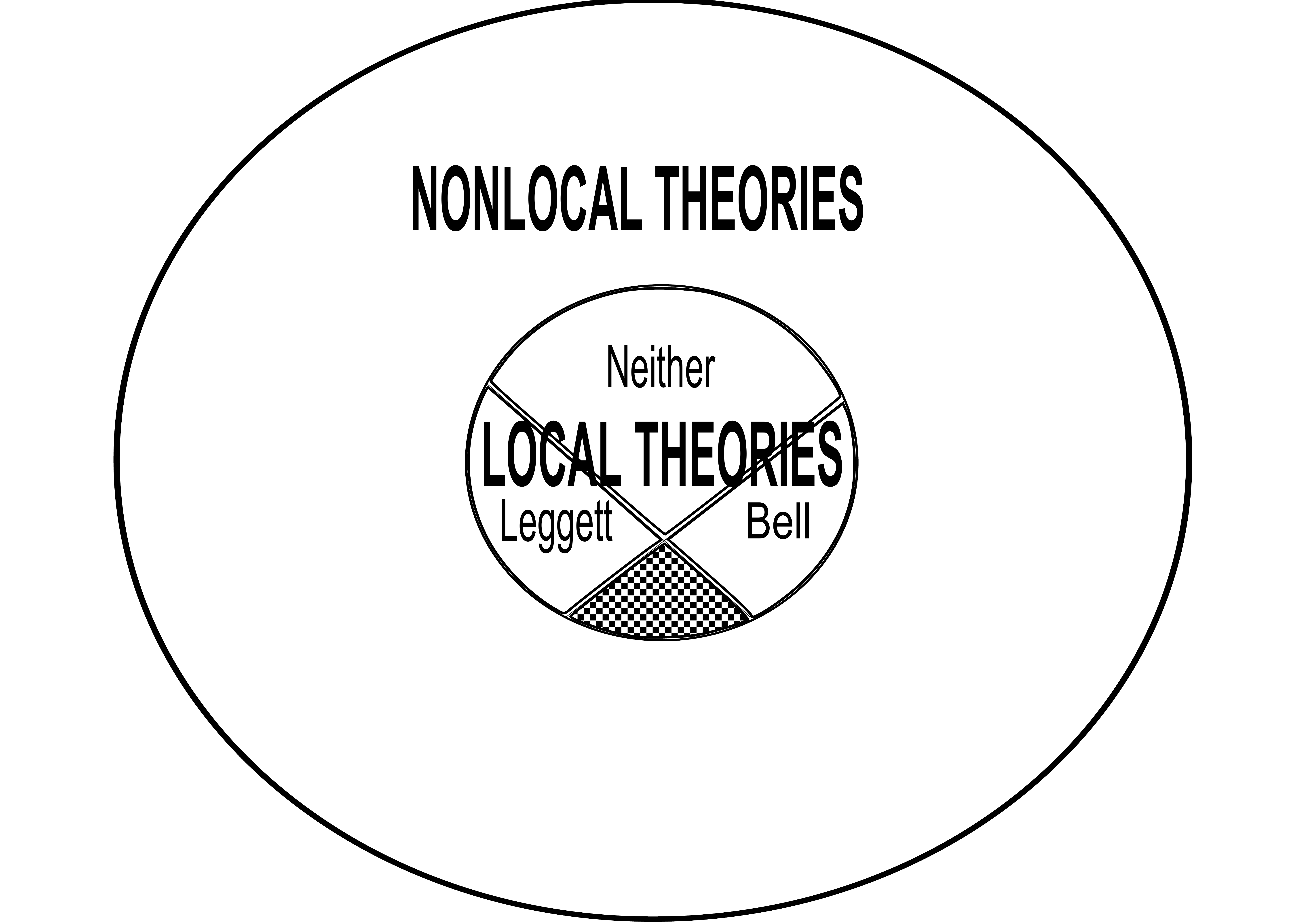}
\caption{\label{fig:venn} A cartoonish depiction of the current situation for hidden-variable models.
Theories satisfying Leggett (i.e. Hypotheses (1)-(3) given by us)
or Bell hypotheses are a subclass of all the possible local theories. They have a nonzero 
intersection represented by the checkered area. Both subclasses are ruled out by experiment. 
There is left a subclass of local theories which do not comply with neither Leggett or Bell hypotheses, and thus 
may violate both inequalities. This is a necessary, but not sufficient, condition to reproduce the current 
experimental evidence.}
\end{figure}
\end{center}
In conclusion, we have demonstrated that the experimental violation of the Leggett inequality rules out 
an important class of stochastic local theories, identified by hypotheses (1)-(3), rather than 
deterministic nonlocal theories, identified by hypotheses (A)-(D).  
Thus the claims that ``nonlocal realism'' was disproved were shown to be unfounded. 
On the other hand, the models ruled out by the experiments do not necessarily obey the property of outcome independence. 
This hypothesis is needed in addition to locality in order to derive the Bell inequality. 
In the derivation of Leggett inequality, this hypothesis is replaced by 
the requirement that a kind of Malus's law applies to the hidden variables. 
Thus the class of models excluded by the violation of Leggett inequality is not contained in the class 
excluded by the violaton of Bell inequality. The union of these classes, however, does not 
exhaust all the possible local hidden-variable theories. 
In a forthcoming paper \cite{DiLorenzo2011b}, the present author will discuss 
a class of local models that do not satisfy neither the outcome independence 
or the compliance with Malus's law hypothesis. These models will be shown to violate both Bell and 
Leggett inequalities.

This work was supported by Funda\c{c}\~{a}o de Amparo \`{a} Pesquisa do 
Estado de Minas Gerais through Process No. APQ-02804-10.

\section*{Appendix}

Equation \eqref{eq:basicineq2} is our starting point. 
Let us study the term 
\begin{equation}\label{eq:upper}
R(\mathbf{a},\mathbf{b})\equiv 
\int dF \left|\mathbf{u}\cdot\mathbf{a}-\mathbf{v}\cdot\mathbf{b} \right| \ .
\end{equation}
The value of the function inside the absolute value is independent of the choice of coordinate system, so its functional 
dependence on the coordinates must change with the coordinate system. 
The same is true for the function $F$.  
In Ref.~\cite{Leggett2003} a fixed coordinate system 
was used, so that the functional dependence of $F$ on its parameters is given. 
We indicate this fixed function in spherical coordinates as 
$F_{\mathcal{S}}(\theta_u,\theta_v,\phi_u,\phi_v)$. 
It is also assumed that $\mathbf{a}_{\mathcal{S}}=(\cos{\phi_a},\sin{\phi_a},0)$ and 
$\mathbf{b}_{\mathcal{S}}=(\cos{\phi_b},\sin{\phi_b},0)$, i.e. the polarizations of the detectors 
vary in the $XY$ plane of the fixed coordinate system. We call 
$\mathbf{p}$ the unit vector along the $Z$ axis in the coordinate system $\mathcal{S}$. 
The vectors $\mathbf{u}$ and $\mathbf{v}$ 
can be written as $\mathbf{u}=\sin{\theta_u}(\cos{\phi_u},\sin{\phi_u},0)+\cos{\theta_u}(0,0,1)$
and $\mathbf{v}=\sin{\theta_v}(\cos{\phi_v},\sin{\phi_v},0)+\cos{\theta_v}(0,0,1)$, 
with $\theta_u\in[0,\pi]$ and $\theta_v\in[0,\pi]$ the angles with the fixed $Z$ axis. 
After defining $\xi\equiv (\phi_a+\phi_b)/2$ and $\phi\equiv \phi_a-\phi_b$, 
it is immediate to check that Eq.~\eqref{eq:upper}
can be rewritten 
\begin{align}
%\nonumber
&R_\mathcal{S}(\phi,\xi)\!= \!\!\int_{0}^{\pi} \!\!\!\! d\theta_u d\theta_v \!\! 
\int_{-\pi}^{\pi} \!\!\!\!\!\! d\phi_u d\phi_v 
%\\
\label{eq:upper2}
F_{\mathcal{S}}(\theta_u,\theta_v,\phi_u,\phi_v) 
N \left|\cos{\left(\frac{\phi_u+\phi_v}{2}+\xi+\delta\right)}\right|.
\end{align}
with $N (\ge 0)$ and $\delta$ defined by 
\begin{align}
N\cos{\delta}=&\left(\sin{\theta_u}-\sin{\theta_v}\right) \cos{\frac{\phi_u-\phi_v-\phi}{2}} \ , \\
N\sin{\delta}=&\left(\sin{\theta_u}+\sin{\theta_v}\right) \sin{\frac{\phi_u-\phi_v-\phi}{2}} \ .
\end{align}
By averaging Eq.~\eqref{eq:basicineq2} 
over the possible values of $\xi$, i.e. rotating $\mathbf{a}$ and $\mathbf{b}$ 
in the fixed $XY$ plane while keeping their relative direction unchanged, 
we obtain
\begin{equation}
-1+L_{\mathbf{p}}(\phi)
\le C_{\mathbf{p}}(\mathbf{a}\cdot \mathbf{b}) \le 
1-R_{\mathbf{p}}(\phi)  \ , 
\end{equation}
where
\begin{align}
\label{eq:upper3}
&R_{\mathbf{p}}(\phi)= \frac{2}{\pi}\int_{0}^{\pi} d\theta_u d\theta_v 
\int_{-\pi}^{\pi} d\phi_u d\phi_v 
F_{\mathcal{S}}(\theta_u,\theta_v,\phi_u,\phi_v) 
N(\theta_u,\theta_v,\phi_u-\phi_v-\phi), \\
\label{eq:lower3}
&L_{\mathbf{p}}(\phi)= \frac{2}{\pi}\int_{0}^{\pi} d\theta_u d\theta_v 
\int_{-\pi}^{\pi} d\phi_u d\phi_v 
F_{\mathcal{S}}(\theta_u,\theta_v,\phi_u,\phi_v) 
N(\theta_u,-\theta_v,\phi_u-\phi_v-\phi) ,
\end{align}
and 
\begin{equation}
C_{\mathbf{p}}(\mathbf{a}\cdot \mathbf{b})\equiv  
\int \frac{d\xi}{2\pi} 
C_{\mathcal{S}}\biggl( \mathbf{a}_{\mathcal{S}}(\xi,\phi),\mathbf{b}_{\mathcal{S}}(\xi,\phi)\biggr) \ .
\end{equation}
Since $N$ depends only on $\phi_u-\phi_v$, $\theta_u$, and $\theta_v$, 
it is convenient to change variables to 
$\chi\equiv \phi_u-\phi_v$ and $\psi=(\phi_u+\phi_v)/2$. 
The two inner integrals in Eq.~\eqref{eq:upper3} then become 
\begin{equation}
\int_{-\pi}^{\pi} d\phi_u d\phi_v (...) = \int_{-2\pi}^{2\pi} d\chi  
\int_{-\pi+|\chi/2|}^{\pi-|\chi/2|}d\psi (...)
\end{equation}
After defining the marginal distribution 
\begin{equation}
\rho_\mathcal{S}(\theta_u,\theta_v,\chi)\equiv
\int_{-\pi+|\chi/2}^{\pi-|\chi/2|} d\psi\ F_{\mathcal{S}}(\theta_u,\theta_v,\psi+\chi/2,\psi-\chi/2)
\ ,
\end{equation}
Eq.~\eqref{eq:upper3} can be rewritten 
\begin{align}
\label{eq:upper4}
R_{\mathbf{p}}(\phi)= \frac{2}{\pi}\int_{0}^{\pi} d\theta_u   d\theta_v 
\int_{-2\pi}^{2\pi} d\chi  
\rho_{\mathcal{S}}(\theta_u,\theta_v,\chi) 
N(\theta_u,\theta_v,\chi-\phi)  \ ,
\end{align}
with 
\begin{equation}\label{eq:kernel}
N^2(\theta_u,\theta_v,\alpha) \!\!=\!\!
\left(\sin{\theta_u}+\sin{\theta_v}\right)^2 \left(\sin{\frac{\alpha}{2}}\right)^2\!+
\left(\sin{\theta_u}-\sin{\theta_v}\right)^2 \left(\cos{\frac{\alpha}{2}}\right)^2
\ .
\end{equation}
Thus, the correlator satisfies
\begin{equation}\label{eq:interm}
-1+L_{\mathbf{p}}(\phi)\le C_{\mathbf{p}}(\phi)\le 1-R_{\mathbf{p}}(\phi) \ .
\end{equation}
By letting the angle between $\mathbf{a}$ and $\mathbf{b}$ vary while keeping both 
vectors in the $XY$ plane, and summing the inequalities, it can be proved, following 
Ref.~\cite{Groblacher2007b}, that 
\begin{equation}\label{eq:xyineq}
 \left|C_{\mathbf{p}}(\phi)+C_{\mathbf{p}}(\phi')\right|\le 
2-\frac{2\sqrt{2}}{\pi}\sin{\left|\frac{\phi-\phi'}{2}\right|} J \ ,
\end{equation}
with 
\begin{equation}\label{eq:ineqcoeff}
J=\int_{0}^{\pi} d\theta_u d\theta_v \mu_\mathcal{S}(\theta_u,\theta_v) 
\sqrt{(\sin{\theta_u})^2+(\sin{\theta_v})^2}\ ,
\end{equation}
and $\mu_\mathcal{S}(\theta_u,\theta_v)=\int d\chi \rho_\mathcal{S}(\theta_u,\theta_v,\chi)$ 
another marginal distribution. 

Now there is a subtle point to be made: After rewriting 
$\mu_{\mathcal{S}}(\theta_u,\theta_v)=\int d\phi_u d\phi_v 
F_{\mathcal{S}}(\theta_u,\theta_v,\phi_u,\phi_v)$, 
Eq.~\eqref{eq:ineqcoeff} can be cast in invariant form, so that  
\begin{equation}\label{eq:invineq}
 \left|C_{\mathbf{p}}(\phi)+C_{\mathbf{p}}(\phi')\right|\le 
2-\frac{2\sqrt{2}}{\pi}\sin{\left|\frac{\phi-\phi'}{2}\right|} \int dF \sqrt{2-(\mathbf{p}\cdot\mathbf{u})^2-(\mathbf{p}\cdot\mathbf{v})^2}
 \ .
\end{equation}
Now, if we consider another unit vector $\mathbf{p}'$ and let $\mathbf{a}$ and $\mathbf{b}$ 
vary in the plane orthogonal to $\mathbf{p}'$, letting the angle 
between them take the same values $\phi$ and $\phi'$, 
the inequality reads 
\begin{equation}\label{eq:invineq2}
 \left|C_{\mathbf{p'}}(\phi)+C_{\mathbf{p'}}(\phi')\right|\le 
2-\frac{2\sqrt{2}}{\pi}\sin{\left|\frac{\phi-\phi'}{2}\right|} \int dF 
\sqrt{2-(\mathbf{p'}\!\cdot\mathbf{u})^2-(\mathbf{p'}\!\cdot\mathbf{v})^2}
 \ .
\end{equation}
By summing Eqs. \eqref{eq:invineq} and \eqref{eq:invineq2}, and noticing that 
$\boldsymbol{[}$see Eqs.~(37-43) of Ref.~\cite{Groblacher2007b} for a proof$\boldsymbol{]}$
\[\sqrt{2-(\mathbf{p}\cdot\mathbf{u})^2-(\mathbf{p}\cdot\mathbf{v})^2}+
\sqrt{2-(\mathbf{p'}\cdot\mathbf{u})^2-(\mathbf{p'}\cdot\mathbf{v})^2}
\ge \sqrt{2} \ ,
\]
when $\mathbf{p}\cdot\mathbf{p}'=0$, we arrive at an inequality  
independent on the knowledge of the measure $dF$, namely 
\begin{equation}\label{eq:leggineq}
 \left|C_{\mathbf{p}}(\phi)+C_{\mathbf{p}}(\phi')\right|+ \left|C_{\mathbf{p'}}(\phi)+C_{\mathbf{p'}}(\phi')\right|\le 
4-\frac{4}{\pi}\sin{\left|\frac{\phi-\phi'}{2}\right|} \ , (\text{for}\ \mathbf{p}\cdot\mathbf{p}'=0) ,
\end{equation}
which is Leggett inequality.

\end{document}